# Voltage-input spintronic oscillator based on competing effect for extended oscillation regions

*Zhifeng Zhu[a], Jiefang Deng, Xuanyao Fong, Gengchiau Liang[b]*

Department of Electrical and Computer Engineering, National University of Singapore, Singapore 117576

The stable precession region in the spintronic oscillator with an in-plane magnetic tunnel junction is very narrow under small external fields, restricting its applications such as for microwave generators. Here we show that this region can be greatly enlarged by introducing competing effects between different torques. Moreover, we observe large-angle precessions at zero external field, which leads to large output powers. We further evaluate the oscillator performance in a voltage-input device, where the circuit area can be minimized and the difficulty of accurate current control can be resolved. The operating voltage window in the proposed device is over 1.23 V, and its frequency can be adjusted from 1.6 to 4.9 GHz. A maximum output power of 0.28 μW is obtained at an energy consumption of 2.2 mW. This study should provide insights for designing voltage-input spintronic oscillators.

# I. INTRODUCTION

As a main functional block in the radio-frequency communication system, the voltage controlled oscillator generates oscillating electrical signals. Traditional ring oscillators and LC tank oscillators occupy large silicon areas (>1000 μm$^2$ [1, 2]), and have limited frequency tunability [3]. To overcome these problems, recently the spin-torque nano-oscillator (STNO) has been proposed as a compact microwave generator [4]. A simple STNO consists of a magnetic tunnel junction (MTJ) driven by dc currents. The polarized electrons transfer angular momentum to the free layer (FL), generating the spin transfer torque (STT). Steady-state magnetization precessions can be obtained by balancing the STT with the Gilbert damping torque [5-7], which induces resistance oscillations in the MTJ due to the tunnel-magnetoresistance (TMR) effect. Since the current is fixed, an alternating voltage signal across the MTJ can be measured. The first STNO is realized in an in-plane MTJ, where an external magnetic field ($\mathbf{H}_{ext}$) is required [6]. Its frequency can be tuned from 12 to 22 GHz with magnetizations precessing in small angles, and the maximum measured output power ($P_o$) is 3.2 nW, which is insufficient for device applications. An enhanced $P_o$ can be obtained at large-angle precessions or out-of-plane precessions, which require large $\mathbf{H}_{ext}$ and STT [6, 8]. To get rid of $\mathbf{H}_{ext}$ and maintain large $P_o$, MTJs with mixed in-plane and perpendicular magnetized layers have been proposed [9-18], but their energy consumptions are also increased due to the large MTJ resistance. Compared to the STT, the spin-orbit torque (SOT) generated by the spin-Hall effect (SHE) [19-21] or the inverse spin galvanic effect (ISGE) [22-24] in the device with a MTJ deposited on top of a heavy metal (HM) layer has higher spin-injection efficiencies [25]. The stable magnetization precessions in this device, namely the spin-Hall nano-oscillator (SHNO), can be obtained by balancing the SOT and damping torque [26-29].

Notably, when $H_{ext}$ is small, both STNOs and SHNOs rely on the delicate balance between the spin torque and damping torque, which makes the devices unfavorable from the following aspects: Firstly, a slightly exceeding spin torque would switch the MTJ to the opposite state, and the oscillating state can only be retrieved by applying an opposite dc current, which complicates the device operations. The resulting input window, where stable oscillations persist, is then very small, e.g., ~ 0.1 mA for $H_{ext}$ < 400 Oe [30]. As a result, the oscillators are vulnerable to perturbations such as thermal fluctuations. In addition, since a precise balance between the spin torque and damping torque is required, the oscillation stability is strongly determined by the quality of the dc current source. While most research has been focused on the physical characteristics of STNOs, the peripheral circuits are less considered [31]. In the CMOS technology, the dc current is obtained using a current mirror [32], which however, not only increases the device area but also causes a scientific problem about how to accurately copy the current from the current source to STNOs under all circumstances [31].

To avoid the problem of accurate current control in current-input devices, in this work, we propose a voltage-input competing spintronic oscillator (VICSO). By introducing opposite effective fields between STT and SOT, we find that the input window is greatly extended, which is attributed to the competing effects between different torques. To verify this proposal, both macrospin and micromagnetic simulations are performed, and similar results are observed. We then simulate the device together with peripheral circuits, and investigate the effects of field-like torque (FLT), thermal fluctuations and material parameters. It is shown that the FLT not only changes the frequency, but also affects dynamic patterns. The thermal fluctuations broaden the linewidth and reduce $P_o$, and dynamic mode transitions between out-of-plane precessions and large-angle precessions are observed when the material parameters are modified.

## II. METHODS

The magnetization dynamics under effective magnetic fields is described by the Landau-Lifshitz-Gilbert-Slonczewski (LLGS) equation,

$$\dot{\mathbf{m}} = -\gamma \mathbf{m} \times \mathbf{H}_{\text{eff}} + \alpha \mathbf{m} \times \dot{\mathbf{m}} - \gamma \mathbf{m} \times (\mathbf{m} \times (\mathbf{H}_{\text{STT}} + \mathbf{H}_{\text{SOT}}))$$
$$-\gamma \mathbf{m} \times (\beta_{\text{STT}} \mathbf{H}_{\text{STT}} + \beta_{\text{SOT}} \mathbf{H}_{\text{SOT}}) \quad , \quad (1)$$

where $\gamma$, $\alpha$, and $\mathbf{m}$ are the gyromagnetic ratio, the damping constant, and the normalized magnetization vector, respectively. The first two terms on the right hand side describe the precession and the Gilbert damping. The effective magnetic field, $\mathbf{H}_{\text{eff}}$, is contributed by the crystalline anisotropy field $\mathbf{H}_{\text{an}}$, the demagnetizing field $\mathbf{H}_{\text{demag}}$, the dipole field $\mathbf{H}_{\text{dipole}}$, the thermal field $\mathbf{H}_{\text{thermal}}$, and $\mathbf{H}_{\text{ext}}$, where $\mathbf{H}_{\text{an}} = 2(K_{\text{bulk}} + K_{\text{i}}/t_{\text{FL}})m_z \hat{\mathbf{z}} / M_{\text{S}}$ with the bulk anisotropy $K_{\text{bulk}} = 2.245 \times 10^5$ J/m³, the interface anisotropy $K_{\text{i}} = 1.286 \times 10^{-3}$ J/m², and $M_{\text{S}} = 1.58$ T extracted from Ref. [33]. $\mathbf{H}_{\text{demag}}$ and $\mathbf{H}_{\text{dipole}}$ are formulated as

$$\mathbf{H}_{\text{demag}} = -(N_x m_x \hat{\mathbf{x}} + N_y m_y \hat{\mathbf{y}} + N_z m_z \hat{\mathbf{z}}), \quad (2)$$

$$\mathbf{H}_{\text{dipole}} = -(D_x m_x \hat{\mathbf{x}} + D_y m_y \hat{\mathbf{y}} + D_z m_z \hat{\mathbf{z}}), \quad (3)$$

where their corresponding tensor components $N_x$, $N_y$, $N_z$ [34], $D_x$, $D_y$, and $D_z$ [35] are analytically calculated. The analytical method for $\mathbf{H}_{\text{demag}}$ is valid for rectangular magnets, and the obtained $\mathbf{H}_{\text{dipole}}$ is verified by comparing with standard numerical results. The thermal fluctuation is included as a random field

$$\mathbf{H}_{\text{thermal}} = N_1(0,u)\hat{\mathbf{x}} + N_2(0,u)\hat{\mathbf{y}} + N_3(0,u)\hat{\mathbf{z}}, \quad (4)$$

where $N_i(0,u)$ is the normal distribution with a zero mean and a standard deviation $u = \sqrt{2k_B T \alpha / (V_{\text{FL}} M_S \gamma (1+\alpha^2) \Delta t)}$ [36] determined by the Boltzmann constant $k_B$, the temperature $T$, the volume of FL $V_{\text{FL}}$, and the duration of thermal fluctuations $\Delta t$. The MTJ resistance in the parallel state, $R_P$, is determined by its dimensions and the resistance-area ($RA$) product (1.5 $\Omega \cdot \mu m^2$ in this work [37]), and the angle and voltage dependences are captured by

$R = R_\mathrm{P} + (R_\mathrm{AP} - R_\mathrm{P})(1 - \cos(\theta))/(2(1 + V_\mathrm{MTJ}/V_\mathrm{h})^2)$ where $\theta$ is the angle between the magnetizations of the FL and PL, $R_\mathrm{AP}$ is the MTJ resistance in the antiparallel (AP) state, and $V_\mathrm{h}$ is the bias voltage giving half *TMR* [38]. $V_\mathrm{h}$ used in the simulation is 0.5 V unless otherwise stated, *TMR* = 150% is extracted from experiments, a large $\alpha$ of 0.025 is used by considering the spin pumping effect [39-41], and the thickness of the MTJ spacer layer is fixed to 1 nm, which is used in the calculation of **H**$_\mathrm{dipole}$.

The last two terms are current induced torques including the STT and SOT. The STT is first studied by Slonczewski [42] and Berger [43] as a damping term to describe the momentum transfer between electron spins and magnetizations, which is formulated as $\mathbf{H}_\mathrm{STT} = \eta \hbar J_\mathrm{STT} \boldsymbol{\sigma}_\mathrm{STT} / (2 e t_\mathrm{FL} M_\mathrm{S})$ with the FL thickness $t_\mathrm{FL}$, the electron charge $e$, the spin polarization $\boldsymbol{\sigma}_\mathrm{STT}$, the reduced Planck constant $\hbar$, and the STT efficiency $\eta = P/[1 + P^2 \cos(\theta)]$ [38] determined by the magnetic layer polarization $P$. The STT also manifests as a FLT, and there are controversies on whether the FLT is a quadratic [44-48], linear [49-51], or more complicated function [52, 53] of the bias voltage. In this study, we simply assume a linear relation between the FLT and the electrical current.

There are also similar debates on the SOT. On one hand, the SHE [19] generates vertical spin currents which transfer the angular momentum to magnetizations similar to that in the STT. On the other hand, spin accumulations [22-24] at the FM/HM interface are exchange coupled with magnetic moments, also producing torques on the FM layer. As an example, the Rashba-Edelstein effect (REE) [22] induces spin accumulations due to the inversion asymmetry and spin orbit coupling [54]. It is shown that both SHE and REE give rise to two types of torques, i.e., the damping like torque (DLT) and the FLT. However, the theory of SHE claims that the FLT is negligible compared to the DLT [55], whereas the FLT is dominant within the theory of REE [22]. Therefore, the SOT in this study is modelled as

$$\mathbf{H}_{\text{SOT}} = \hbar \mathbf{J}_{\text{SOT}} / (2et_{\text{FL}} M_S), \tag{5}$$

$$\mathbf{J}_{\text{SOT}} = \theta_{\text{SH}} \boldsymbol{\sigma} \times \mathbf{J}_C (1 - \text{sech}(t_{\text{HM}} / \lambda_{\text{sf}})), \tag{6}$$

where $t_{\text{HM}}$ is the HM thickness, $\lambda_{\text{sf}}$ is the spin-flip length, and the hyperbolic secant function is originated from the vertical spin drift and diffusion [55].

## III. RESULTS AND DISCUSSION

### A. Operation mechanism

We first study the device operation by analyzing the effective fields. Fig. 1 illustrates the effective fields for both P and AP states, where $\mathbf{J}_{\text{SOT}}^{-y}$ and $\mathbf{J}_{\text{STT}}^{+y}$ denote the respective current densities of the SOT and STT with the polarizations showing in the superscripts. The generation of $\mathbf{J}_{\text{SOT}}^{-y}$ under positive $V_{\text{DD,SOT}}$ requires a HM with negative $\theta_{\text{SH}}$, such as W or Ta [56]. To illustrate the competing effects, the STT is controlled separately using $V_{\text{DD,STT}}$. The PL magnetization is fixed in the $-\hat{\mathbf{y}}$ direction, which can be stabilized by an antiferromagnetic (AFM) exchange bias layer. The FL magnetization is designed to switch back and forth between $m_y = 1$ and $m_y = -1$ by balancing different fields. An intuitive understanding of device operations can be obtained by comparing effective fields between the P and AP states. In the AP state, $\mathbf{H}_{\text{demag,AP}}$ and $\mathbf{H}_{\text{SOT,AP}}$ are opposite to $\mathbf{H}_{\text{STT,AP}}$ and $\mathbf{H}_{\text{ext}}$. Once the fields in the $-\hat{\mathbf{y}}$ direction dominate, the magnetization is switched to the P state, accompanied by the reversal of $\mathbf{H}_{\text{demag}}$. The back switching to the AP state can be achieved by ensuring larger fields in the $+\hat{\mathbf{y}}$ direction. As a result, the MTJ will be switched back and forth between the AP and P states once the abovementioned two conditions are satisfied, which is summarized in the bottom of Fig. 1, and a large oscillation region can be obtained based on this competing effect. Another difference with STNOs is the voltage input. Since $\mathbf{I}_{\text{MTJ}}$ in STNOs is fixed, the oscillating voltage can be analytically obtained by assuming an angle dependent MTJ resistance $R_{\text{MTJ}}$ [57]. In contrast, both $\mathbf{I}_{\text{MTJ}}$ and $R_{\text{MTJ}}$ in the VICSO are oscillating although the supply voltage ($V_{\text{DD}}$) is fixed. Therefore, the output voltage is not a simple function, and a numerical method such as the distribution model described in Appendix C is required to capture the oscillator characteristics.

**B. Oscillation dynamics, effect of field like torque and thermal fluctuations**

To verify device operations, in this section, the oscillator is simulated without any FLT at zero $T$, and detailed studies including both effects will be discussed later. Firstly, we study magnetization dynamics without the competing effects, which can be realized by setting $V_{\text{DD,STT}} = 0$. In this case, the effective fields include $\mathbf{H}_{\text{ext}}$, $\mathbf{H}_{\text{SOT}}$, $\mathbf{H}_{\text{demag}}$, and $\mathbf{H}_{\text{an}}$, and the resulting phase diagram showing in Fig. 2(a) is similar to that in conventional STNOs [6]. Note that there are regions with small-angle precessions between the transitions from AP to P, which are not visible in Fig. 2(a) since they are very narrow. The oscillation patterns of the device driven only by $V_{\text{DD,SOT}}$ is identical to that from the STNO driven by the constant current source due to the similar expression of the effective field (i.e., $\mathbf{H}_{\text{STT}} = \eta \hbar J_{\text{STT}} \boldsymbol{\sigma}_{\text{STT}} / (2 e t_{\text{FL}} M_{\text{S}})$ and $\mathbf{H}_{\text{SOT}} = \hbar \mathbf{J}_{\text{SOT}} / (2 e t_{\text{FL}} M_{\text{S}})$, in which both $J_{\text{STT}}$ and $J_{\text{SOT}}$ are constant). Interestingly, when the STNO is driven by the constant voltage source, we have observed additional oscillation patterns, which are stabilized by the negative feedback between $J_{\text{STT}}$ and the STT efficiency (refer to Appendix A for detailed discussions). Moreover, when the competing effects are introduced by setting $V_{\text{DD,STT}} = V_{\text{DD,SOT}}$, the oscillation region at small $\mathbf{H}_{\text{ext}}$ is further enlarged as shown in Fig. 2(b), and this result is qualitatively reproduced in micromagnetic simulations discussed in Appendix B. In addition, large-angle precessions also appear at zero $\mathbf{H}_{\text{ext}}$, which, in current-input STNOs and SHNOs, can only be obtained under large $\mathbf{H}_{ext}$ and spin torques. At large $V_{DD}$, both $\mathbf{H}_{STT}$ and $\mathbf{H}_{SOT}$ are sufficient to switch the FL to the opposite state, distorting the magnetization trajectory into a large-angle precession. This is similar to the large-angle precession observed in STNOs, where $\mathbf{H}_{ext}$ is opposite to $\mathbf{H}_{STT}$ and both are sufficient for switching [6]. As a result, the STT in our device behaves partly as $\mathbf{H}_{ext}$, which also explains why there is an extended oscillation region at small $\mathbf{H}_{ext}$. Therefore, the appearance of large-angle precessions at zero $\mathbf{H}_{ext}$ is attributed to both the negative feedback formed in the voltage-input device as well as the competing effects between different torques.

Once the device operations are verified, we then evaluate the oscillator performance by simulating the device together with peripheral circuits as shown in Fig. 3(a), where $\mathbf{H}_{dipole}$ generated by the PL is used to replace $\mathbf{H}_{ext}$. The three resistances can be used as additional controls to adjust current distributions to fulfill the oscillation conditions showing in Fig. 1. Once $\mathbf{m}_{FL}$ is excited, $R_{MTJ}$ oscillates due to the TMR effect, which changes the current distribution and induces alternating voltage signals on $R_3$. It is worth noting that the drop of electrical potential across the HM layer induces nonuniform distributions of $\mathbf{J}_{SOT}$ and $\mathbf{J}_{STT}$, which is captured using a distributed circuit model (see Appendix C).

Referring to the phase diagram in Fig. 2(b), $\mathbf{H}_{dipole}$ generated by the PL is 436 Oe, which corresponds to $H/H_C = 0.82$, and the applied $V_{DD}$ is 1.09V corresponding to $V/V_C = 4$. As a result, $\mathbf{m}_{FL}$ precesses around the $z$ axis with a tilted angle [see Fig. 3(b)], with $m_y$ ranges from $-1$ to 0.6. The resulting oscillating output voltage ($V_o$) has a peak to peak value of 7 mV [see Fig. 3(c)], which confirms the functionality of the VICSO.

To have reasonable evaluations of the device performance, the FLT and thermal fluctuations have to be included. According to theoretical calculations and experimental measurements, the FLT in the STT is much smaller than the DLT, with $\beta_{STT}$ ranging from 0 to 0.3 [46]. In contrast, due to the complex physical origins in the SOT, there are still controversies about whether the FLT or DLT dominates in the FM/HM bilayer, where $\beta_{SOT}$ ranges from 0 to 3 [55, 58, 59]. Therefore, we sweep both $\beta_{STT}$ and $\beta_{SOT}$ in this study. As shown in Fig. 3(d), a blueshift in frequency is observed when $\beta_{SOT}$ is increased, whereas the frequency is a decreasing function of $\beta_{STT}$, e.g., for $\beta_{STT} = 0.15$, the frequency increases from 2.33 to 3.37 GHz when $\beta_{SOT}$ is increased from 0 to 3; for $\beta_{SOT} = 1.1$, the frequency decreases from 2.69 to 2.65 GHz when $\beta_{STT}$ is increased from 0 to 0.3. The decreasing function of $\beta_{STT}$ is attributed to the cancellation between the STT and SOT. Since $\beta_{SOT}$ is larger than $\beta_{STT}$ in most regions, the frequency is mainly determined by $\beta_{SOT}$, and the increase in $\beta_{STT}$ reduces the net FLT, resulting in a lower frequency. In addition, there are certain regions without any oscillations, indicating that the FLT not only changes the frequency, but also has a qualitative effect on the oscillation dynamics. The zero-oscillation regions appear when $\beta_{STT}$ and $\beta_{SOT}$ have similar magnitudes. The alternating $\mathbf{H}_{STT}$ induced by the oscillation produces a changing FLT, which might contribute to the zero-oscillation regions. In other regions where $\beta_{SOT}$ is much larger than $\beta_{STT}$, the effect of the alternating FLT is negligible, and oscillations are always sustained. As shown in Eq. (4), $\mathbf{H}_{\text{thermal}}$ is determined by its duration $\Delta t$. We choose $\Delta t = 5$ ps, which is much smaller than the precession period. The effect of thermal fluctuations is discussed in Appendix D, where a linewidth broadening and a reduction in the output voltage are observed.

In the following discussions, both FLT and thermal fluctuations are included. The thermal fluctuations are modelled at $T = 300$ K, and $\beta_{STT} = 0.15$ and $\beta_{SOT} = 1.1$ are selected, which is away from the zero-oscillation regions. In addition, these FLT coefficients agree with the theories that the DLT is dominant in the STT, whereas the FLT is dominant in the SOT.

## C. Voltage tunability

As shown in Fig. 4, we further investigate the voltage tunability of the proposed VICSO. Firstly, the stable oscillation region is identified by sweeping $V_{DD}$. Based on the parameters used in this study, the stable oscillation exists for $V_{DD} > 0.73$ V [see Fig. 4 (a)], and it can be tuned by modifying material parameters. For $V_{DD} < 0.73$ V, no oscillations are excited due to the small spin torques, and the magnetization stays in the initial state. $V_{DD} > 2$ V is not investigated considering the dielectric breakdown of the MTJ [60]. The oscillation frequency is determined by performing the fast Fourier transform (FFT) on $V_O$ for over 600 periods. As an example, the FFT spectrum at $V_{DD}$ = 1.09 is shown as the inset in Fig. 4(a), where a fundamental frequency ($f_0$) of 2.69 GHz and a second harmonic at $2f_0$ are observed. The emergence of the second harmonic indicates that the precession velocity is not a constant in one period, resulting in a distorted trajectory away from the perfect circle. From the efficiency point of view, the high order harmonics are preferred to be weak to concentrate the energy at the fundamental frequency. By repeating this procedure, the frequencies at different $V_{DD}$ are determined, and a monotonic increase from 1.4 to 4.9 GHz is observed [see Fig. 4(a)], which is attributed to the enhanced spin torque. Next, the effect of $V_{DD}$ on power consumption ($P_c$) and $P_o$ are studied. $P_c$ is computed as the sum of power consumed in the MTJ, HM, and resistors. As shown in Fig. 4(b), $P_c$ (blue circles) increases with $V_{DD}$ in a monotonic way, contributed by the increased power dissipation in each part, and a minimum $P_c$ of 1.3 mW is identified at $V_{DD}$ = 0.77 V. $P_o$ is evaluated by $P_o = 10\log_{10}(P_{ave}/1\text{mW})$ where $P_{ave}$ is the averaged output power on $R_3$. As illustrated in in Fig. 4(b) using yellow squares, $P_o$ firstly increases with a peak of $-39.2$ dBm at $V_{DD}$ = 1.33 V, and it then saturates. For $V_{DD}$ below 1.33 V, the peak to peak value of $m_y$ varies in a small range. Therefore, $P_o$ increases for larger $V_{DD}$. However, the peak to peak value of $m_y$ is a decrease function of $V_{DD}$ for $V_{DD} > 1.33$

V, which compensates the increase of $V_{DD}$, resulting in $P_o$ with small variations. Combining both trajectories, an optimum operation point appears at $V_{DD} = 1.33$ V with $P_o/P_c = 3\%$.

**D. Effect of magnetic, electrical, and dimensional parameters**

Finally, we explore the effects of magnetic ($\alpha$, TMR, P, $M_S$, $\theta_{SH}$), electrical ($R_3$), and dimensional parameters (MTJ length). Fig. 5 shows the frequency and $P_o$ as a function of the parametric ratio $R_i = u_i/u_0$, where the subscript $i$ specifies the parameter type, and $u_i$ and $u_0$ denote the value and reference value, respectively. For $R_3$ (and $P$, not shown here), the frequency is a monotonic decrease function of $R$, whereas it is an increase function for $\theta_{SH}$ (and TMR, not shown here). Both trends can be obtained by analyzing the effective fields, e.g., the increase of $\theta_{SH}$ enhances $\mathbf{H}_{SOT}$, which speeds up the precession of magnetization. In contrast, the increase of $R_3$ reduces the current flowing through the HM, resulting in smaller $\mathbf{H}_{SOT}$ and hence a smaller frequency. However, this simple analysis only applies when the oscillation patterns remain the same. As shown in the curve of $\alpha$ (and $M_S$, not shown here), the frequency is not a monotonic function of $R$, and a transition from out-of-plane precessions to large-angle precessions is observed when $R_\alpha$ across 1.24. As shown in Fig. 5(b), there is an opposite trend between the frequency and $P_o$, e.g., $P_o$ as a function of $\theta_{SH}$ decreases monotonically. The high frequency oscillation requires a small precession angle to reduce the total trajectory, leading to small magnetization swings and hence small $P_o$. In addition, $P_o$ as a function of $\alpha$ saturates at large $R_i$, which is because the change of $m_y$ swing is very small at large $R_i$. Comparing $P_o$ under different parameters, the maximum $P_o = -35.6$ dBm is obtained at $R_3 = 201$ $\Omega$.

## IV. CONCLUSION

In conclusion, we have proposed a voltage-input spintronic oscillator based on the competing effects between different torques. By performing both macrospin and micromagnetic simulations, we show that the competing effects greatly enlarge the stable oscillation regions, especially at small external fields. In addition, large-angle precessions are observed at zero external field, which permits large output powers. The proposed voltage-input device is then simulated together with the peripheral circuits to verify its operation. As a result, a wide operating voltage window over 1.23 V and a tunable microwave frequency from 1.6 to 4.9 GHz are achieved. After systematically studying the parametric effects, an oscillation mode transition from out-of-plane precessions to large-angle precessions is observed, and an optimum operation point is identified with the largest $P_o/P_c$. The large input window and rich oscillation dynamics can be useful in the applications of spintronics oscillators such as microwave generators or beyond CMOS computing.

**Corresponding Authors**

[a)]a0132576@u.nus.edu, [b)]elelg@nus.edu.sg

**ACKNOWLEDGEMENTS**

We would like to thank Dr. Gaurav Gupta for his contribution to the theoretical understanding. This work at the National University of Singapore was supported by CRP award no. NRF-CRP12-2013-01.

**Appendix A: Voltage-input versus current-input oscillators**

The conventional STNO is mainly studied under the constant current source, where only small-angle precession is permitted [see Fig. 3(a) of Ref. [6]]. Interestingly, we found that additional oscillation patterns are allowed when the STNO is driven by the constant voltage source. The devices under investigation are illustrated in Fig. 6. In the current-input device, small-angle oscillation exists at $I_c$ = 0.658 mA and magnetization switching happens for a slightly increased current. In the voltage-input device, the small-angle precession also appears when the damping is balanced by the spin torque. However, an increased $V_{DD}$ leads to large-angle precessions (e.g., $V_{DD}$ = 0.26 V) before the magnetization is switched. The appearance of additional precession patterns can be understood by noticing that $J_{STT}$ and the STT efficiency, $\eta = P/[1+P^2\cos(\theta)]$ which is larger in the AP state, form a negative feedback, i.e., the device in Fig. 6(b) is initially in the P state, the STT tends to switch $m_y$ towards +1, resulting in a smaller $J_{STT}$, which is compensated by the increasing $\eta$. Therefore, a metastable state is maintained due to this negative feedback. We have also compared three different configurations for the device shown in Fig. 1, (i) only $V_{DD,STT}$ is applied, (ii) only $V_{DD,SOT}$ is applied, and (iii) both $V_{DD,STT}$ and $V_{DD,SOT}$ are applied, in which we have observed qualitatively different frequency trend and oscillation patterns. Therefore, the oscillation in the VICSO is unlikely originated from the synchronization [61, 62] between the STT- and SOT-induced oscillations.

In addition, it is necessary to compare the VICSO with the three-terminal SHNO reported in Ref. [27]. Despite the similarity in the device structure, the fixed current source is used in the SHNO, in which the large-angle precession under zero $\mathbf{H}_{ext}$ cannot be achieved. In addition, $I_{MTJ}$ in the SHNO mainly modifies the anisotropy through the voltage-controlled magnetic anisotropy effect (VCMA), whereas our simulation without the VCMA effect provides more clear oscillation patterns under the competing effect.

**Appendix B: Micromagnetic simulations**

The micromagnetic simulation is carried out in the structure shown in Fig. 1. The simulation is performed in the Mumax3 [63] which has been modified to enable the fix voltage simulation. The modified source code can be found in the seeder branch at

https://github.com/xfong/3. $M_S$, $\alpha$, $\theta_{SH}$, $K_{bulk}$, $K_i$, $\rho_{HM}$, and the FL size are the same as that in the macrospin simulation described in the main text. The exchange stiffness $A_{ex}$ = 13 pJ/m, and the FL is discretized into 50×90×1 cells. For the STT simulations, the polarizations of both FL and PL are set to 0.4, and the Slonczewski Λ parameters of both PL and FL are set to 2. In the SOT simulations, the polarization of the free layer and Slonczewski Λ parameter are set to 1, and we assume that $\mathbf{J}_{SOT}$ is independent of $t_{HM}$. All the micromagnetic simulations are carried out without any FLT at $T$ = 0.

As shown in Fig. 7(a), the device without competing effects has an oscillation window from 0.05 V to 0.12 V at zero field. By introducing the competing effects, an enlarged oscillation window from 0.2 V to 0.47 V is observed [see Fig. 7(b)]. The expansion of oscillations due to the competing effects agrees with the macrospin simulations shown in Fig. 2.

**Appendix C: Distributed Circuit Model**

It is worth noting that both $\mathbf{J}_{SOT}^{-y}$ and $\mathbf{J}_{SOT}^{+y}$, as depicted in Fig. 3(a), are spatially nonuniform because of the drop in electrical potentials along the $x$ direction. To capture this, we adopt a distributed model, which segments the HM and MTJ into $N$ parts as depicted in Fig. 8(a). Each MTJ segment is modeled as a resistor whose value is determined by its dimensions and $RA$ product. To get the correct branch currents and voltage potentials, this distributed model is iteratively solved. As $N$ is increased, the results become more accurate, and a consistent solution will be obtained for a sufficient $N$, which is 50 in our study. After that, the average current in the MTJ and HM ($I_{M,i}$ and $I_{HM,i}$) are used to determine $\mathbf{J}_{STT}$ and $\mathbf{J}_{SOT}$, which are used to compute the effective fields. The LLGS equation is then solved to get the magnetization dynamics, and then the resistances are calculated again. These simulation procedures, summarized in Fig. 8(b), are repeated to get the time evolution of magnetizations.

**Appendix D: Effect of thermal fluctuations**

The frequency spectrums for the oscillations with and without thermal fluctuations are calculated at $V_{DD}$ = 1.09 V. As shown in Fig. 9(a), a fundamental frequency of 2.69 GHz and a second harmonic at 4.456 GHz are observed, where the linewidth is very narrow. By including the thermal fluctuations at $T$ = 300 K, the fundamental frequency remains the same at 2.71 GHz, whereas the full-width-at-half-maximum (FWHM) linewidth is greatly expanded to 320 MHz.

**Figures**

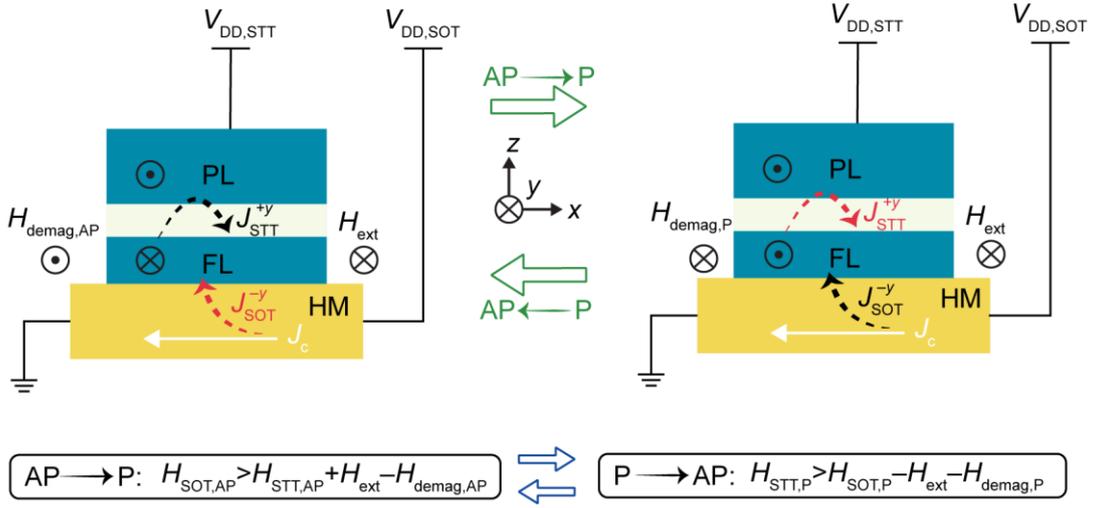

Figure 1. Schematic representations of the device structure and effective fields at the AP and P states. In the P state, $\mathbf{H}_{ext}$, $\mathbf{H}_{demag}$, and $\mathbf{J}_{STT}$ are in the $+\hat{\mathbf{y}}$ direction, which are opposite to $\mathbf{J}_{SOT}$. Once the combined fields in the $+\hat{\mathbf{y}}$ direction dominate, the magnetization is switched to the AP state, resulting in reversed $\mathbf{H}_{demag}$. The magnetization switching is accompanied by the change of current distributions such as $\mathbf{J}_{STT}$ and $\mathbf{J}_{SOT}$. The device parameters are chosen that the combined fields in the $-\hat{\mathbf{y}}$ direction dominate in the AP state, and then the magnetization is switched back to the P state. By fulfilling the abovementioned two conditions summarized in the bottom of the figure, a back and forth switching between the P and AP state can be obtained, resulting in magnetization oscillations. The dimensions of MTJ are 50 nm × 90 nm with the FL (PL) thickness of 3 (4) nm, and the FL polarization is 0.4. The size of HM is 100 nm × 100 nm × 5 nm, with the resistivity $\rho_{HM} = 200 \times 10^{-8} \Omega \cdot m$ [56], $\lambda_{sf} = 5$ nm, and $|\theta_{SH}| = 0.1$.

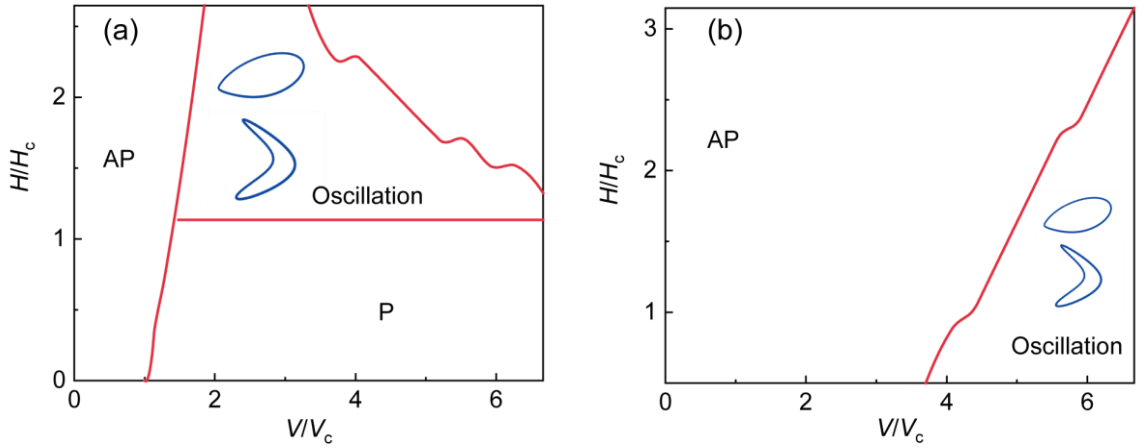

**Figure 2**. Phase diagrams of the device (a) without and (b) with competing effects between **H**$_{STT}$ and **H**$_{SOT}$. $H_C$ = 529 Oe and $V_C$ = 0.27 V are defined as the critical field and the STT switching voltage, respectively. The voltage induced MTJ resistance reduction is not included in both devices. The red lines separates different dynamic regions. The oscillation trajectories are illustrated using blue lines.

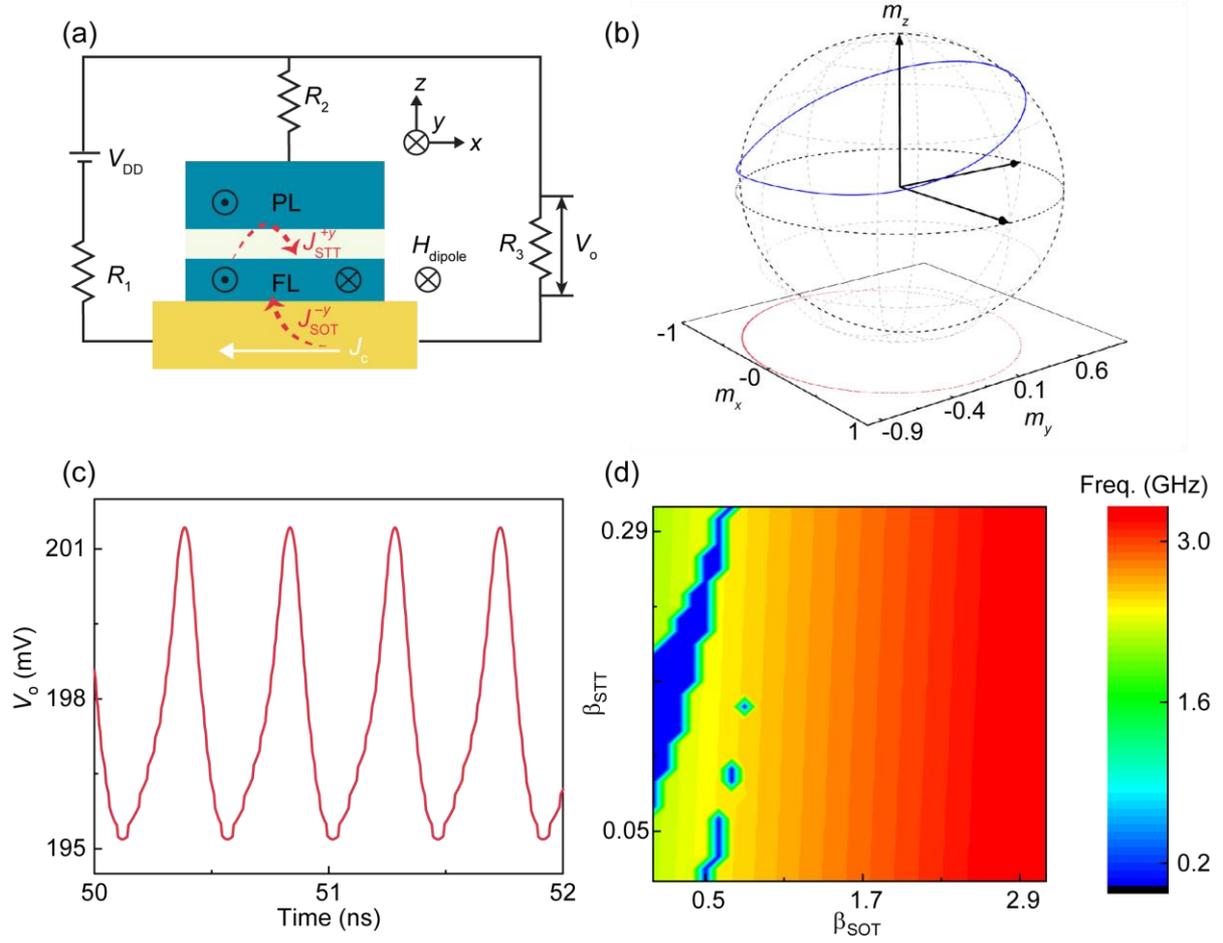

**Figure 3**. (a) The complete device diagram, including a three terminal MTJ/HM structure, the voltage source, and the three resistors $R_1$, $R_2$, $R_3$. Referring to the device structure illustrated in Fig. 1, the STT and SOT are controlled using a single voltage $V_{DD}$, $\mathbf{H}_{ext}$ is replaced by $\mathbf{H}_{dipole}$ generated by the PL, and magnetization oscillations are translated into output voltage signals on $R_3$. (b) Magnetization trajectories (blue solid line) with the projection on the $m_x$-$m_y$ plane (red dotted line) at $V_{DD}$ = 1.09 V. The magnetization rotates around the $\hat{\mathbf{z}}$ axis with a tilted angle. (c) The time evolutions of the alternating output voltage, where the peak to peak value is 7 mV. There is an incubation time of 10 ns after applying $V_{DD}$, which is removed during the frequency calculation. (d) Oscillation frequency as a function of $\beta_{STT}$ and $\beta_{SOT}$ at $V_{DD}$ = 1.09 V. The frequency is shown in a linear color scale from 0 (blue) to 3.4 GHz (red).

The thermal fluctuations are included at $T = 300$ K. $\beta_{SOT}$ is swept from 0 to 3, and $\beta_{STT}$ is swept from 0 to 0.3. No oscillations are observed in the blue regions.

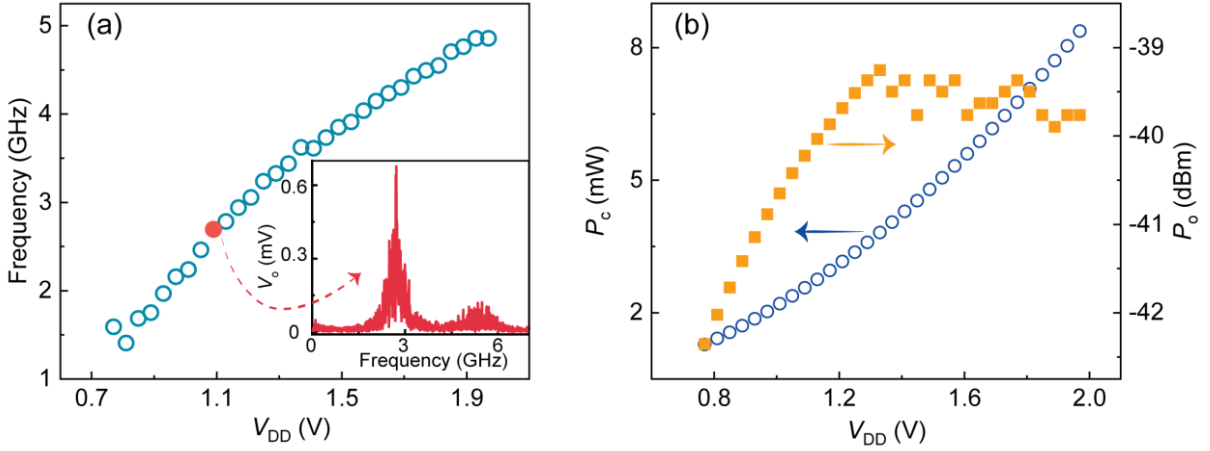

**Figure 4**. (a) Frequency as a function of $V_{DD}$, with the inset showing the spectrum diagram at $V_{DD} = 1.09$ V (red dot). For $V_{DD} < 0.73$ V, no oscillation is observed. $V_{DD} > 2$ V is not investigated considering the dielectric breakdown in the MTJ. (b) $V_{DD}$ dependence of the power consumption (circles) and output power (squares). The peak to peak value of $m_y$ decreases for $V_{DD} > 1.33$ V. The maximum output power is $-39.2$ dBm at $V_{DD} = 1.33$ V.

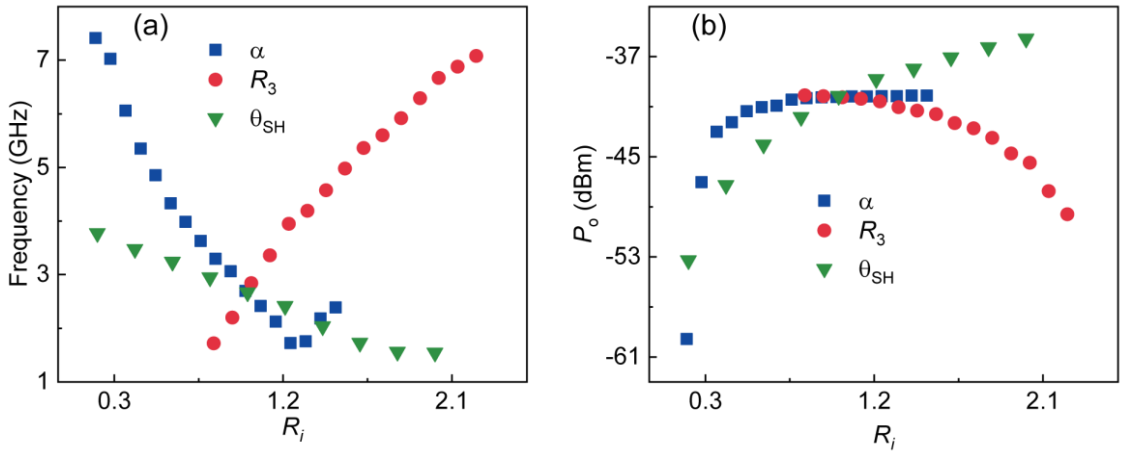

**Figure 5**. (a) Frequency and (b) output power versus the parametric ratio $R_i$ defined as $R_i = u_i/u_0$, where the subscript $i$ specifies the parameter type, and $u_i$ and $u_0$ denote the value and reference value, respectively. i.e., $R_{30} = 100\ \Omega$, $\alpha_0 = 0.025$, $\theta_{SH0} = 0.1$. The trajectory of $\alpha$ is not a monotonic function of $R_i$, accompanied by the transition from out-of-plane precessions to large-angle precessions when $R_\alpha$ across 1.24. All oscillation patterns of $R_3$ and $\theta_{SH}$ are out-of-plane precessions.

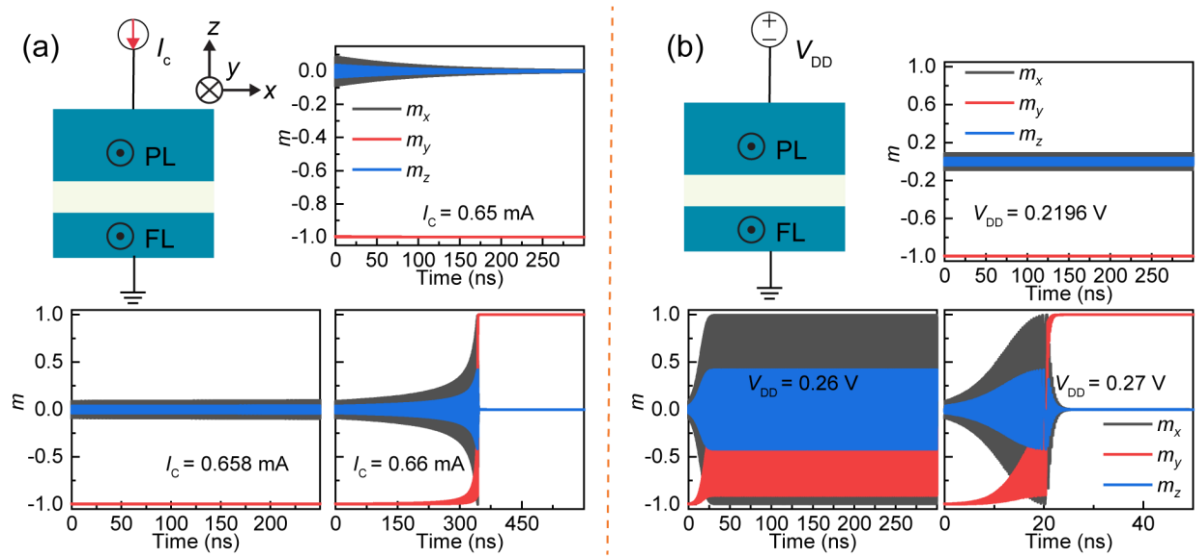

**Figure 6**. (a) STNO driven by the constant current. Only small-angle precession appears for $I_c$ between 0.658 and 0.66 mA. The MTJ size is the same with that in the VICSO. (b) STNO driven by the constant voltage source. Small-angle precession appears at $V_{DD} = 0.2196$ V, and large-angle precession appears for $V_{DD}$ between 0.22 and 0.27 V. All simulations are performed without $\mathbf{H}_{ext}$, FLT, and thermal fluctuations.

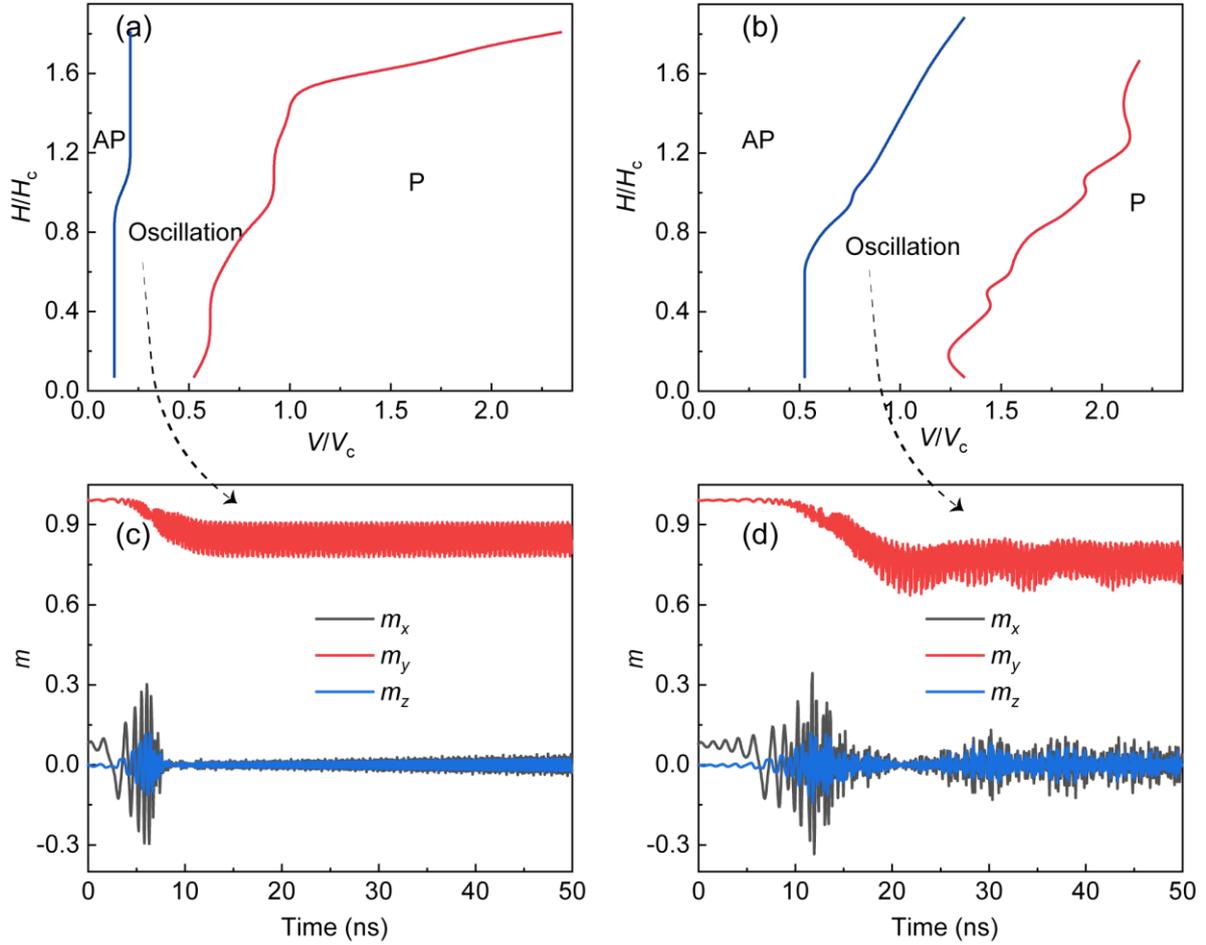

**Figure 7**. Phase diagrams from micromagnetics for the devices (a) without and (b) with the competing effects. $H_c$ = 415 Oe and $V_c$ = 0.38 V are defined as the critical field and the STT switching voltage, respectively. Without the competing effects, the stable precession window at zero field ranges from $0.13V_c$ to $0.32V_c$, whereas it spans from $0.53V_c$ to $1.24V_c$ after introducing the competing effects. (c) Magnetization dynamics at $V_{DD}$ = 0.11 V and $H_{ext}$ = 240 Oe under SOT and $H_{ext}$. (d) Magnetization dynamics at $V_{DD}$ = 0.41 V and $H_{ext}$ = 240 Oe under SOT, STT, and $H_{ext}$.

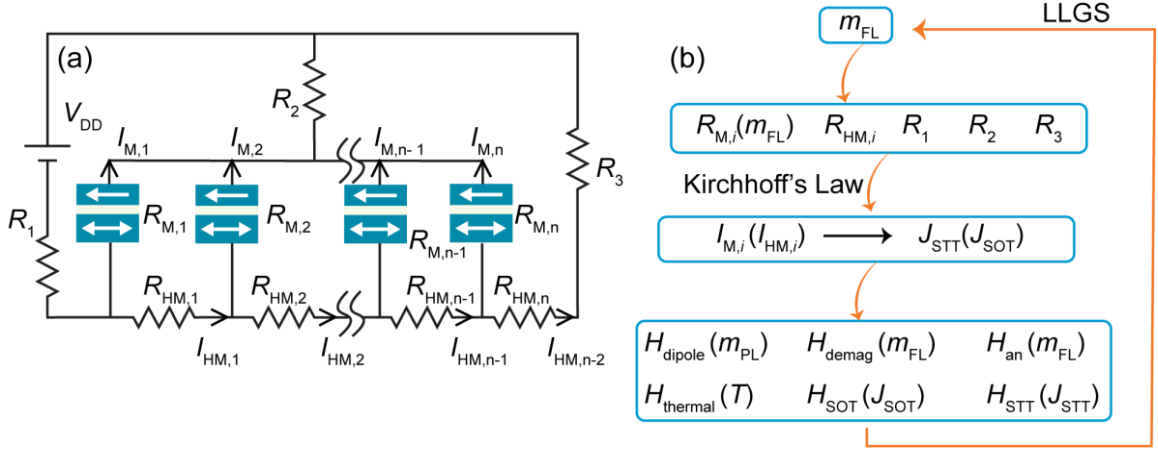

**Figure 8**. (a) The distributed circuit model for the simulation of nonuniform currents and voltage distributions inside the HM and MTJ. The HM and MTJ are divided into $N$ segmental resistors. $\mathbf{I}_{HM,n}$ and $\mathbf{I}_{M,n}$ are the branch currents flowing through the HM and the MTJ, respectively. (b) The computation workflow with parameter dependencies denoted in parentheses, e.g., $R_{M,i}$ depends on $\mathbf{m}_{FL}$. The simulation starts from the resistance computation, followed by the calculation of spin currents using the distributed model. Next, all effective fields are determined, which are input to the LLGS equation to simulate the magnetization dynamics. The equilibrium magnetization is then used to calculate the resistance again.

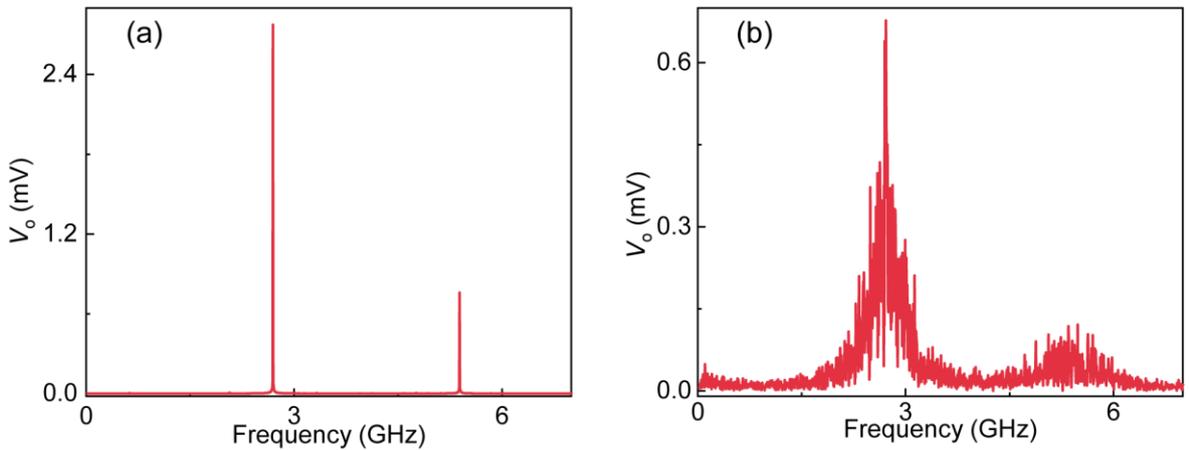

**Figure 9**. Spectrum diagrams of the output voltage at (a) $T = 0$ and (b) $T = 300$. $V_{DD}$ in both cases are 1.09 V. The fundamental frequency in the device without thermal fluctuations is 2.69 GHz with negligible full-width-at-half-maximum (FWHM) linewidth, whereas a similar frequency of 2.71 GHz and a large FWHM of 320 MHz are observed after introducing thermal fluctuations.